# Fabrication of a low-cost and high-resolution papercraft smartphone spectrometer

Young-Gu Ju*

*Department of Physics Education, Kyungpook National University, 80 Daehakro, Bukgu, Daegu, 41566, Korea*
*Corresponding author:ygju@knu.ac.kr*

We demonstrated the fabrication of a low-cost and high-resolution papercraft smartphone spectrometer and characterized its performance by recording spectra from gas discharge lamps. The optical design and a lab-made narrow slit used in the fabrication led to fine images of the slit on the image sensor, resulting in high spectral resolution. The spectral resolution of the fabricated papercraft smartphone spectrometer was measured to be 0.5 nm, which is similar to that of the best smartphone spectrometer reported thus far. Extending the exposure time of the phone's camera revealed the fine structure of a spectrum with high sensitivity. The build cost of the papercraft smartphone spectrometer was less than $3. We demonstrated that the papercraft smartphone spectrometer is a low-cost device that can record spectra with high resolution and high sensitivity.

## I. INTRODUCTION

Since their introduction, smartphones have changed peoples' way of life by combining numerous functions of traditional electronic devices such as phones, computers, cameras, voice and video recorders, and computer networks. Among these technologies, cameras have advanced to the extent that the capabilities smartphone cameras are now comparable to those of traditional digital single-lens reflex cameras in terms of resolution, number of pixels, numerical aperture, autofocusing, and manual controls. The rapid progress in smartphone cameras has opened a path for the development of new portable devices based on smartphones, such as smartphone microscopes and colorimetric devices [1, 2]. A smartphone fluorescence microscope can function as a point-of-care (POC) diagnostic platform because many biodetection methods are based on fluorescence probing [3]. In addition to the attention these devices have attracted, further attention has been devoted to the development of smartphone spectrometers because a few external optical components can transform a smartphone into a high-resolution spectrometer at low cost for various POC applications [4-9].

Smartphone spectrometers have often been fabricated by students as an extracurricular activity. Instructions for fabricating smartphone spectrometers are easily available on the Internet [10]. The housings of most of the smartphone spectrometers for educational use are made of paper, and a digital versatile disc (DVD) is typically used to separate light into its component wavelengths. Paper and DVDs are used mainly to reduce costs because of the limited budgets for such extracurricular activities. Compared with the smartphone spectrometers fabricated in the lab for POC purposes, papercraft smartphone spectrometers (PCSSs) demonstrate lower resolution and sensitivity, as detailed elsewhere [9]. However, if a PCSS was optically well designed with replacement core components, it could potentially provide high resolution and sensitivity while maintaining the low cost of the original papercraft device.

In this paper, we propose a method to fabricate a low-cost, high-resolution papercraft smartphone spectrometer (HRPCSS) and demonstrate its performance by recording the spectra of gas discharge lamps. This work may be useful in the development of an inexpensive smartphone spectrometer for POC or other remote sensing applications.

## II. DESIGN AND FABRICATION

The HRPCSS proposed in this paper differs from conventional PCSSs in two main aspects. A conventional PCSS has a slit made by cutting paper and a grating made from a DVD. Although this method of fabricating the core components is most economical, it also hinders further improvement of device performance.

The use of a paper-cut slit limits the minimum width of the slit. Because the image of the slit is formed on the image sensor of the smartphone and determines the resolution, a narrower slit enables higher resolution, whereas a wider slit enables higher sensitivity or a stronger signal. If the distance between the slit and the smartphone camera is 80 mm, then the magnification is approximately 3/80 because of the effective focal length of a smartphone camera is typically ~3 mm. The 100 μm-wide slit becomes a ~3.8 μm-wide image on the image sensor, which covers approximately four pixels if the pixel size is assumed to be approximately 1 μm. The diffraction limit of the lens also adds more dimension to the size of the geometrical image. When the f/# of the lens is 1, the spot size due to the diffraction limit is approximately 1 μm for visible wavelengths [11]. The width of the slit image on the image sensor becomes approximately 5 μm when both the geometric image and the diffraction limit are considered. In this way, the slit width determines the resolution of the smartphone spectrometer. However, the slit width of a paper cut is greater than 1 mm in most cases. Thus, the resolution of the conventional PCSS is ten times worse than that of a spectrometer with a 100 μm slit. We therefore replaced the paper-cut slit with a more precise but still inexpensive one. In fact, this replacement was investigated in our previous research on a 3D-printed smartphone spectrometer [9]. In the present paper, we provide a more detailed description of the fabrication process for easier duplication of the experiment.

The second improvement was to the grating. A DVD is used as a grating in the conventional PCSS, whereas the one used in the present study was a 1000 lines/mm Rainbow Symphony diffraction grating slide, which is a transmission-type grating that can currently be purchased from Internet shopping sites for approximately US$1. These gratings are produced in large volume with good quality, and they perform better than a DVD without sacrificing the cost advantage. Compared with the reflection-blazed grating used for the 3D printed smartphone spectrometer in our previous research [9], the Rainbow Symphony transmission grating has inferior energy efficiency. The blazed grating concentrates the light energy into the first order, whereas the Rainbow Symphony transmission grating distributes the light energy into other modes such as the zeroth, first, and second modes; less than 30% of the energy is transferred into the first mode, which we use for signal measurements. The light energy transferred into the other modes can generate noise and weaken the signal of the first-order diffraction because of lowered efficiency. Despite the inferior performance of the transmission grating used in the present experiments, it offers a substantial cost advantage over the blazed grating, which is approximately US$70. Therefore, the Rainbow Symphony transmission grating is suitable for use in the HRPCSS in terms of cost and adequate performance.

The assembly view of the HRPCSS is illustrated in Fig. 1. The narrow slit and holographic transmission grating are placed at the ends of the paper housing. The grating side of the housing is sloped to direct the first-order diffraction beam in the direction normal to the surface or smartphone, which places the first-order image in the middle of the field of view. The distance between the slit and the grating or the smartphone camera is approximately 80 mm, which enables the smartphone camera to image the slit by adjusting its focus. When the distance is less than 50 mm, adjusting the focus and obtaining a clear image becomes difficult without an additional external lens.

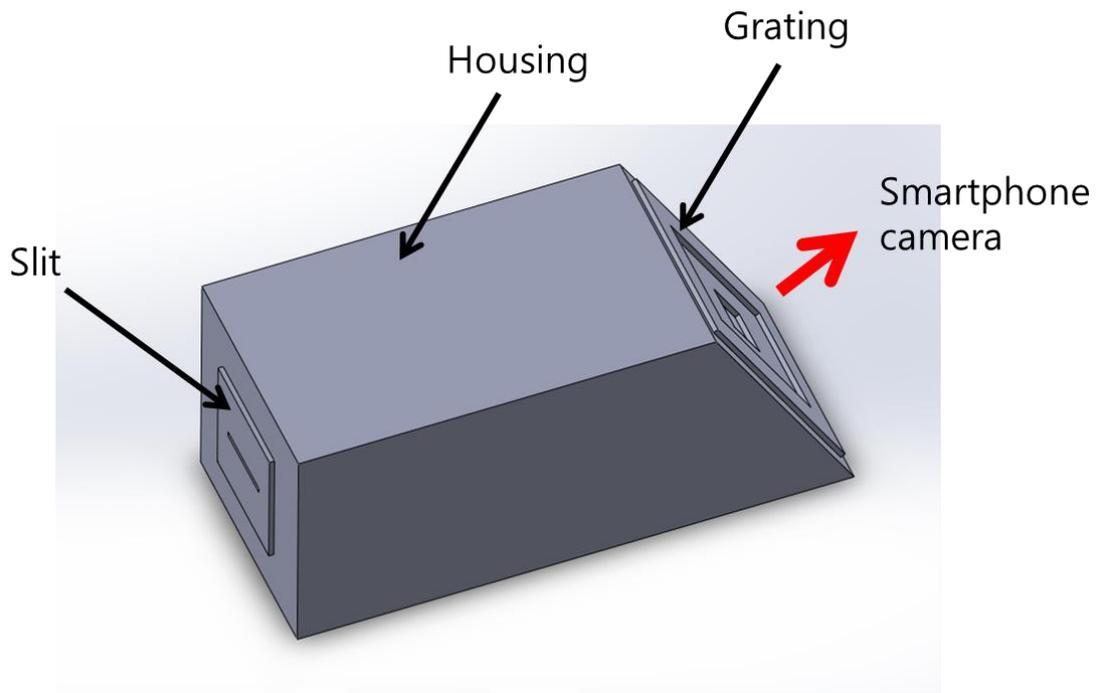

Fig. 1. Assembly view of the papercraft smartphone spectrometer

Prior to describing the construction of the paper folded housing, we first describe the fabrication of the narrow slit. The slit used in the previous study was also made by the doctor-blade method used in the present work [12]. A more detailed description of how to create the narrow slit is provided in Fig. 2. The process begins with applying two pieces of tape to a glass slide with a certain gap between the pieces of tape. Silver paint is then applied to the gap and spread with another slide glass, which functions as a blade. Removal of the tape leaves a silver paint coating with a thickness equal to that of the tape. The finished coating is dried either by being placed in an oven for ~5 min at 90 °C or standing at room temperature for 1 day. After the drying process, the slit is made by drawing a line on the coating with a cutter knife. In this process, the width of the slit depends on which side of the blade is used. In general, the front edge gives a width of ~20 μm and the false edge gives a width of ~100 μm. For intermediate slit widths, we recommend using sand paper to blunt the edge of the knife, which will lead to a wider slit width. In this study, a 100 μm slit was used.

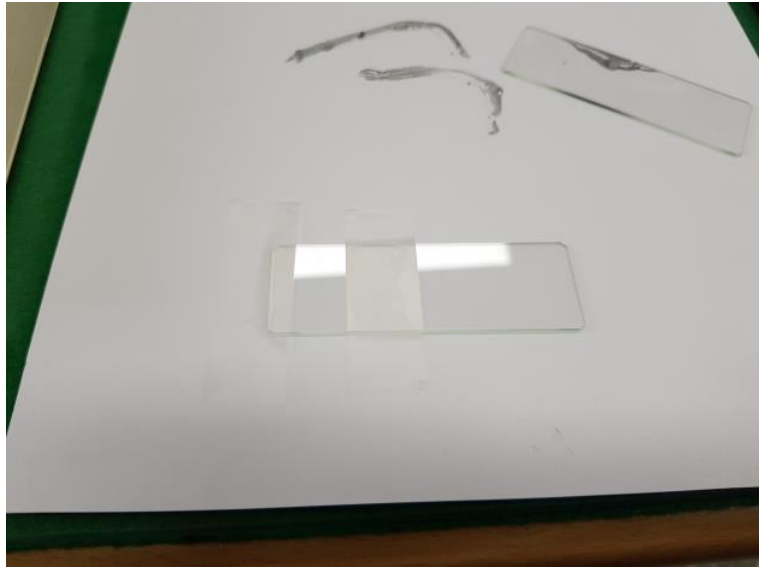
(a)

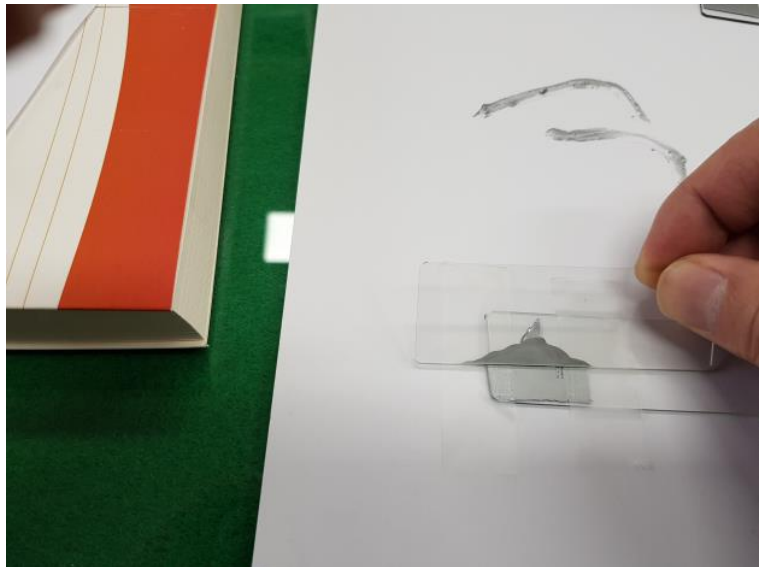
(b)

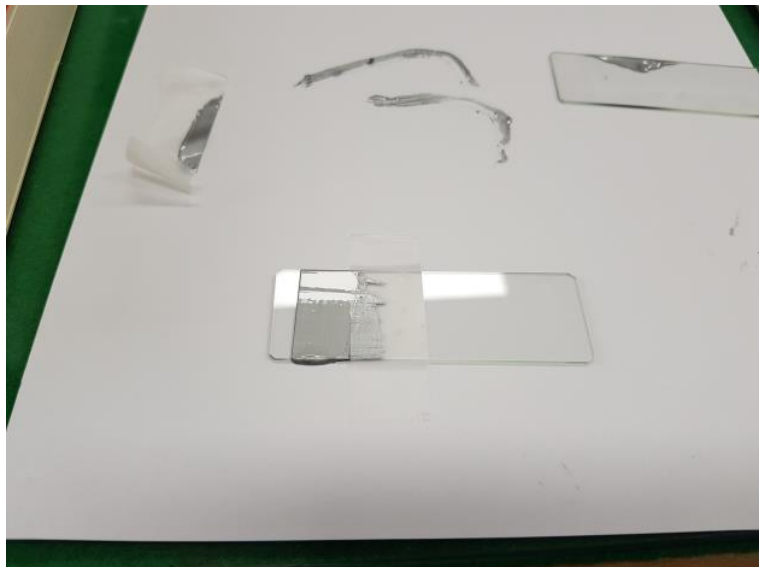
(c)

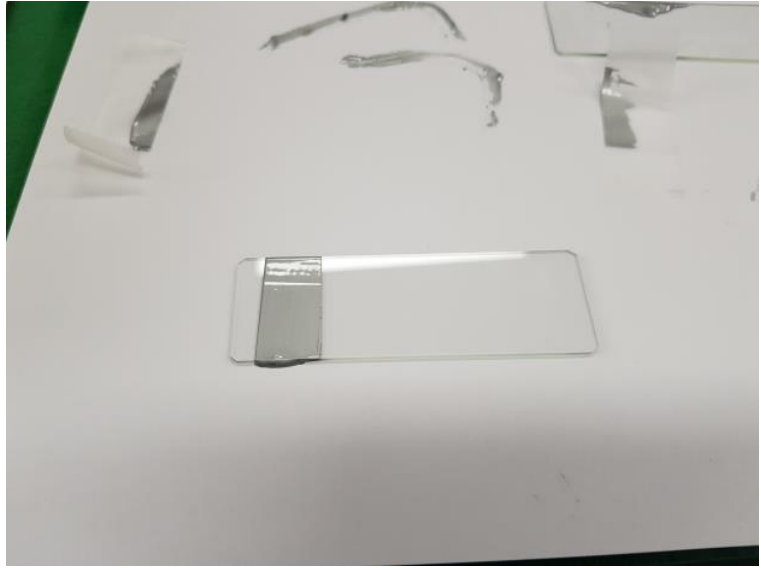
(d)

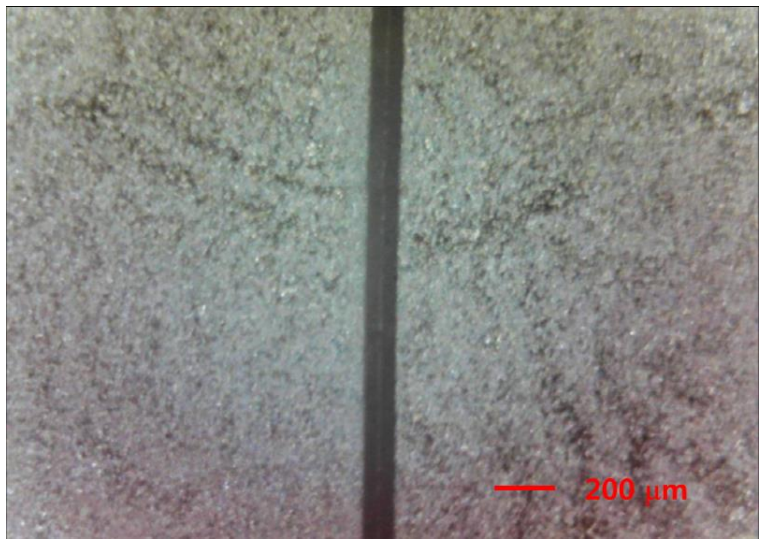
(e)

Fig. 2. Process of making a narrow slit by the doctor-blade method: (a) Tape pieces are applied to a glass slide with a gap between them; (b) silver paint is applied to the gap and then spread with another glass slide. (c) After the paint has been spread, (d) the coating of silver paint remains after the tape is removed. (e) Microscope image of the fabricated slit.

The housing was fabricated by folding paper. Although this paper cutting and folding can be carried out with a cutter knife as usual, we used a 40 W CO2 laser cutter to fabricate numerous smartphone spectrometers with ease and accuracy. The paper was 1 mm-thick cardboard paper that cost less than US$0.20 per piece. The advantage of papercrafting using a laser cutter over 3D printing is a much greater process speed and lower cost. Cutting the edge and engraving the folding edge with a laser requires only a few minutes, whereas a 3D printer typically takes a few hours to print a housing of the same size. The folding edge was created by laser-engraving mode, where the laser beam was moved at higher speed so that the cutting depth was too shallow for complete cutting but adequate for folding. The finished laser-cut paper is shown in Fig. 3(a). The assembly after folding is shown in Fig. 3(b). The structure was fixed by taping the opened edges. After the housing was fixed, the slit and grating were taped to the sides of the housing as shown in Fig. 3(c) and (d), respectively. The assembled HRPCSS was fixed to a smartphone with a tripod, as shown in Fig. 3(e), which is important for mechanical stability during the measurement, especially those involving a long exposure time.

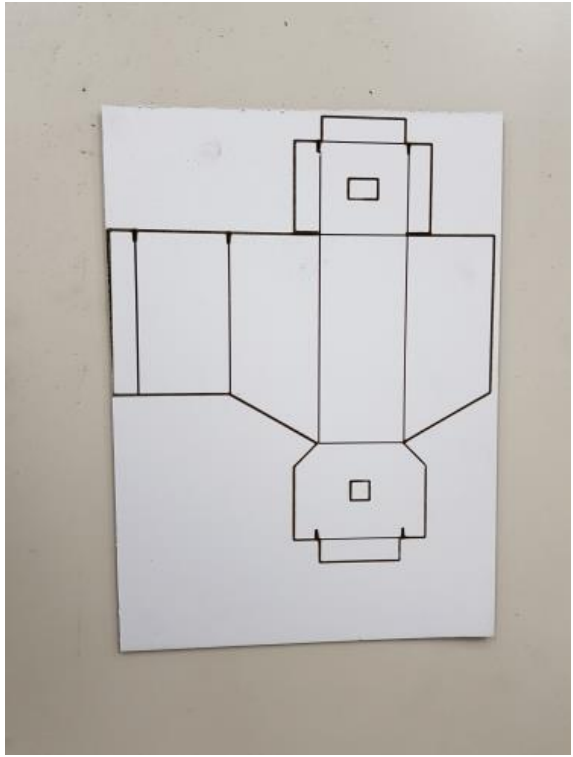

(a)

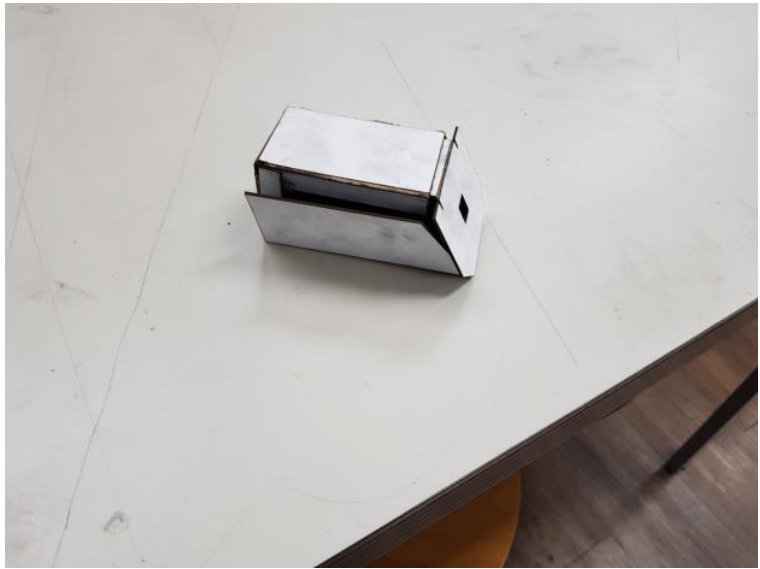

(b)

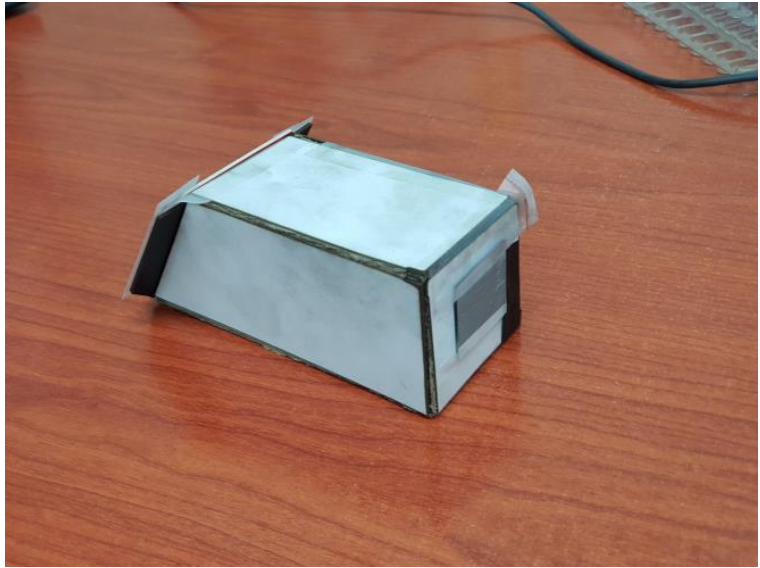

(c)

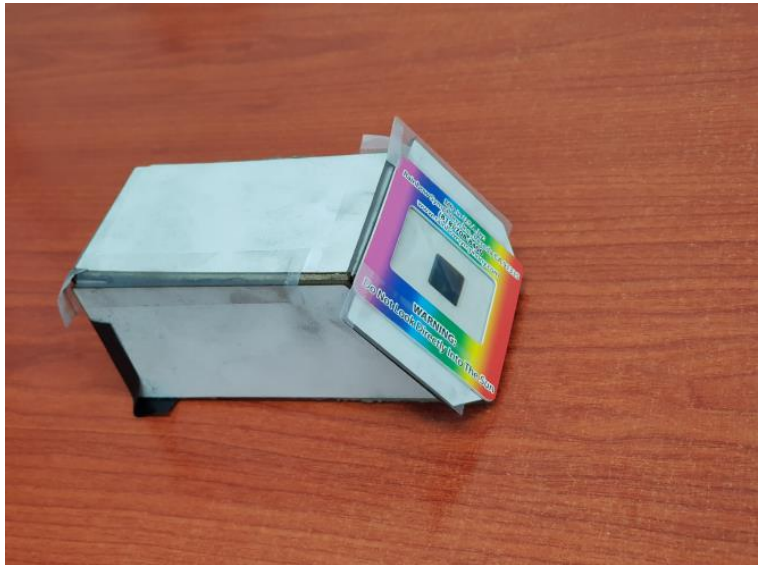

(d)

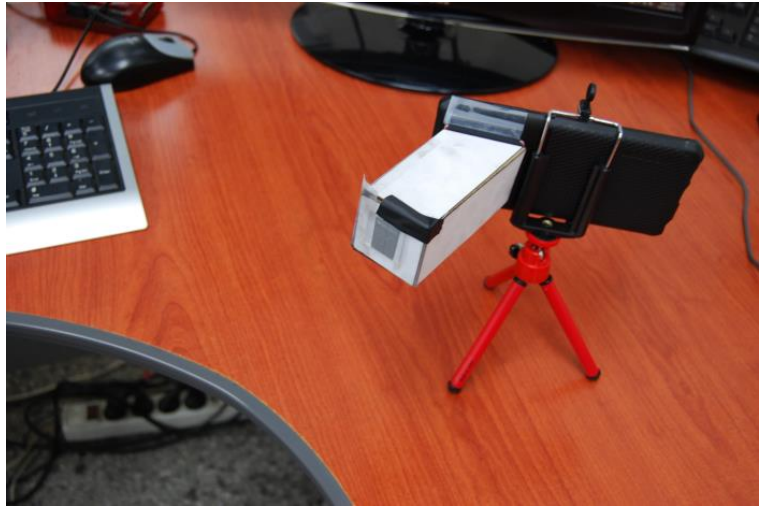

(e)

Fig. 3. Process of fabricating the HRPCSS: (a) The paper cut for the housing after laser cutting and engraving; (b) the papercraft housing after folding; (c) the slit taped to the housing; (d) the grating taped to the housing; (e) the HRPCSS attached to a smartphone (Galaxy Note 8) mounted onto a tripod.

## III. RESULTS AND DISCUSSION

To characterize the performance of the HRPCSS, we used it to record line spectra of gas discharge lamps. Three spectra from a Hg gas lamp are presented in Fig. 4; the three spectra differ with respect to exposure time. With a short exposure time, the four main spectral lines appear near 436 nm, 549.9 nm, 581.6 nm, and 583.8 nm, as shown in Fig. 4(a). As the exposure time increases, finer lines appear (Fig. 4(b) and (c)), whereas the main lines become saturated and broad. The spectrum recorded with an exposure of as long as 10 s was recorded using the manual mode of the smartphone camera. The manual mode control enables adjustment of the focus, ISO value, exposure time, etc. The splitting of the line near 582 nm into two lines demonstrates the improved resolution of the HRPCSS; these lines are not usually resolved by a conventional PCSS.

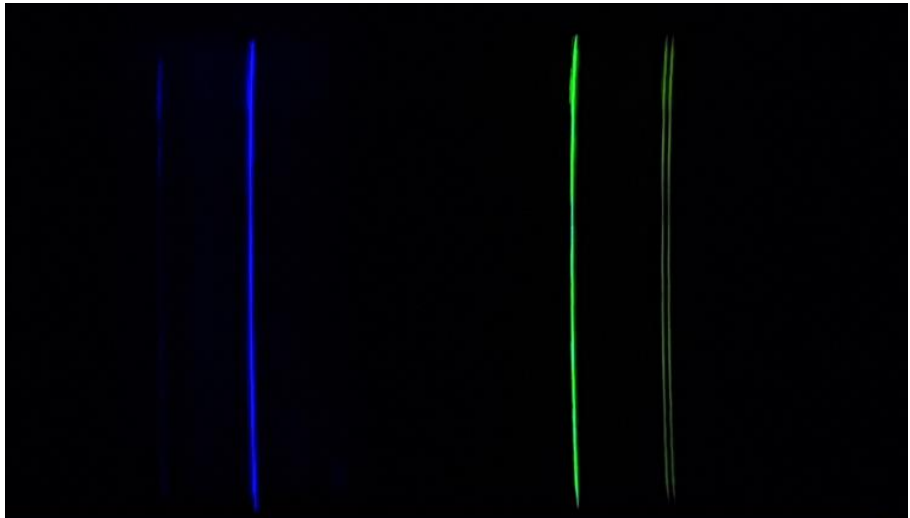
(a)

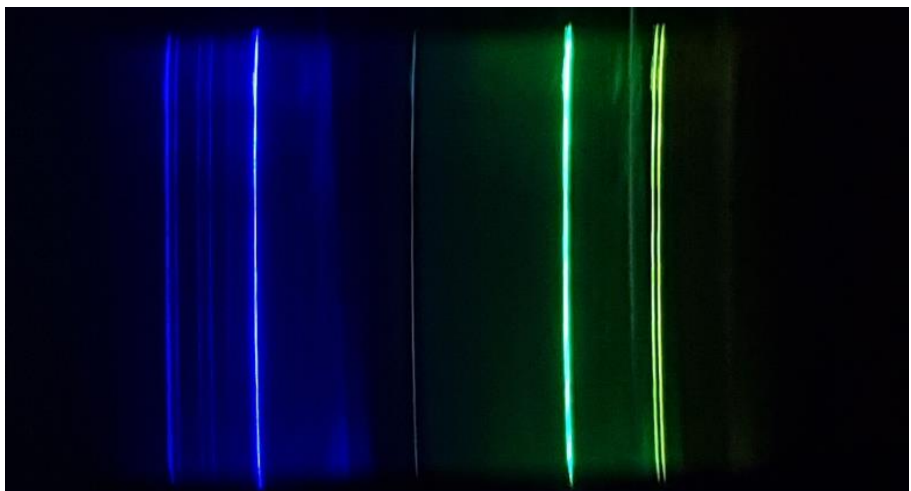
(b)

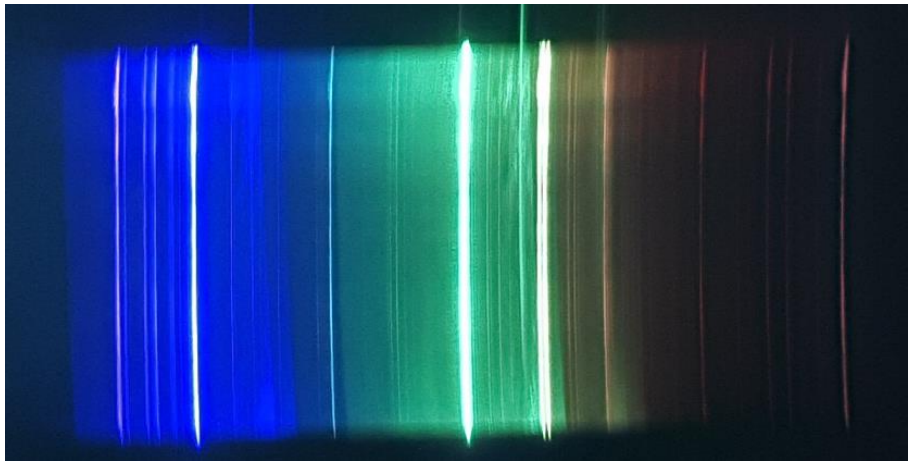

(c)

Fig. 4. Hg-lamp emission spectra recorded by the HRPCSS with exposure times of (a) 1/30 s, (b) 1/2 s, and (c) 10 s

Fig. 5(a) and (b) shows the spectra of Ne and Ar lamps, respectively. Like the spectra of the Hg lamp, resolution is improved and weak lines are revealed with long exposure times.

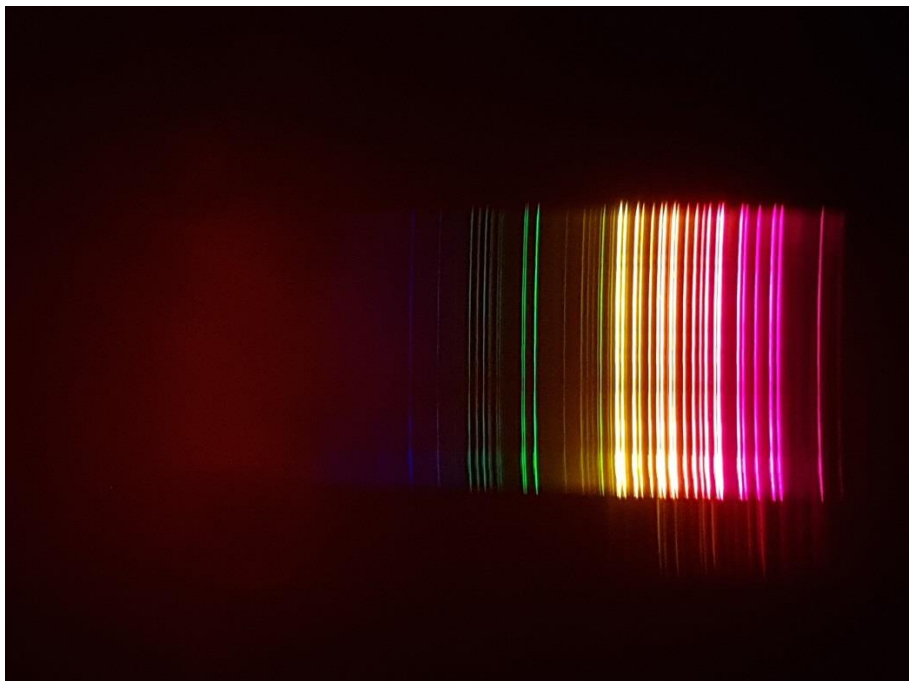

(a)

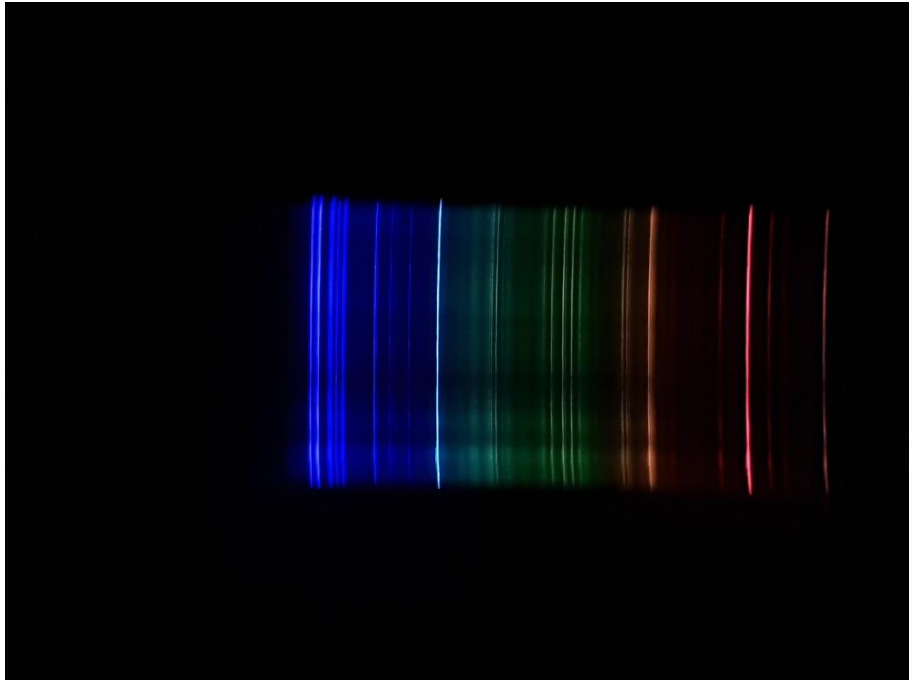

(b)

Fig. 5. The spectra recorded by the HRPCSS from (a) a Ne lamp with exposure of 1/2 s and (b) an Ar lamp with exposure of 10 s

Digitizing the pixel intensity in the acquired spectral image across the spectral lines enables quantification of the spectral intensity at each pixel. The image processing for profiling the cross-section was performed using the ImageJ software [13]. The calibration of the wavelength converts the pixel position into a wavelength by matching the dominant peaks with the wavelengths measured by another calibrated spectrometer. In the case of the Hg lamp, the wavelengths 436.4 nm, 551.1 nm, 582.1 nm, and 584.4 nm were used for calibration. In this way, the intensities of the spectral lines for Hg, Ar, and Ne as a function of wavelength were plotted as shown in Fig. 6.

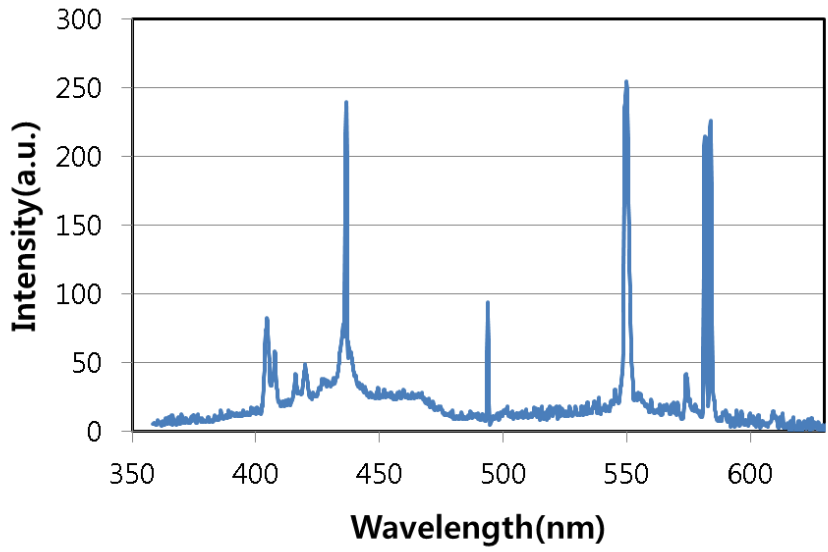

(a)

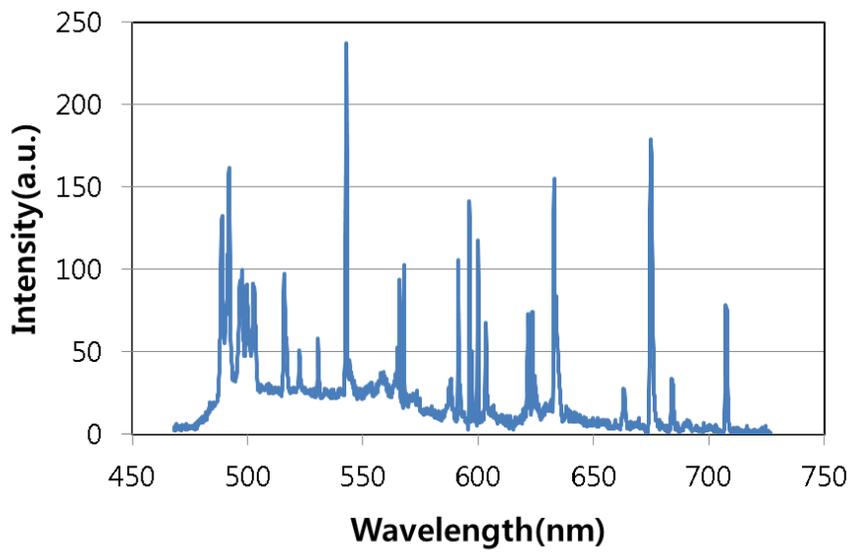

(b)

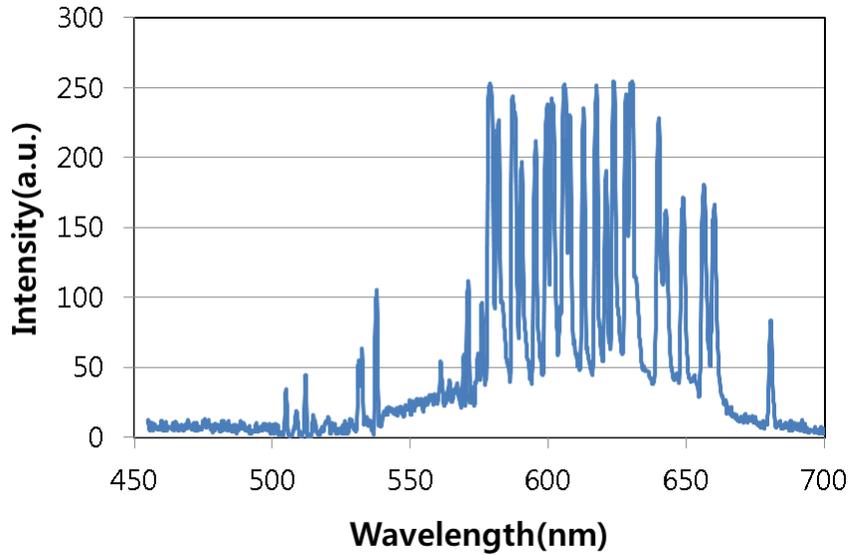

(c)

Fig. 6. Spectral intensity plots obtained from the spectral image as a function of wavelength: (a) Hg, (b) Ar, and (c) Ne.

We experimentally determined the resolution of the spectrometer by measuring the finest linewidth in the spectrum. The finest line is that at ~567.8 nm in the Ar spectrum. Its linewidth is 4.5 pixels and 0.43 nm. The measured resolution is close to the theoretical limit of 4.8 pixels within the measurement error mentioned in the previous section. Because the measured resolution cannot exceed the theoretical limit, the reasonable resolution is 4.8 pixels and 0.46 nm, which is approximately 0.5 nm when the measurement error is considered. The resolution of 0.5 nm is better than those of other smartphone spectrometers [4-6] and is close to the 0.4 nm resolution achieved with a 3D printed smartphone spectrometer in our previous study [9]. The fine resolution of the HRPCSS is surprising given that the PCSS consists of paper and very low-cost optical components.

The HRPCSS can be fabricated for less than $3. The grating and cardboard paper cost less than $2. The microscope slide glass, silver paint, and tape cost less than $1 per unit. The

total cost of the unit is less than $3 in terms of materials. This price is less than those of the smartphone spectrometers reported in Refs. [4-6]. The low fabrication cost of the HRPCSS represents a substantial advantage in many applications such as disposable sensors for POC and outdoor activity programs for education.

Before closing the discussion, we address the sensitivity of the HRPCSS. The performance metric of HRPCSS includes not only the resolution but also the sensitivity, which is the amplitude of the spectrum signal for a light source with a given radiance at the slit. The sensitivity can be increased easily by increasing the ISO value and exposure time of the camera under manual mode. However, increasing the signal by electronic amplification also increases the noise. The optical noise may arise from stray light and the low efficiency of the grating. A picture recorded with long exposure time is displayed in Fig. 7. It shows the zeroth-order, first-order, and the second-order diffraction in the same image. The zeroth-order image is much brighter than the first-order image, which means that more incident light energy is directed to the zeroth-order emission than to the first-order emission, which contributes to the signal we measure. The energy directed to the zeroth order not only reduces the signal but also increases noise by scattering light inside the housing, which increases the background signal. The other source of noise is observed in the zeroth-order image in Fig. 7. A long exposure time reveals that the rectangular area around the slit is also bright. If the silver paint on the glass blocked the light completely, this area would be dark. Therefore, the partial transmission of the area covered with silver paint can be a source of noise during long exposures because the image of the slit becomes the spectral image in the first-order diffraction. An easy method to increase the signal-to-noise ratio is to use a wider slit in place of the narrow slit at the expense of resolution. In some sensor applications, the sensitivity can be more important than the resolution, in which case this method would be useful.

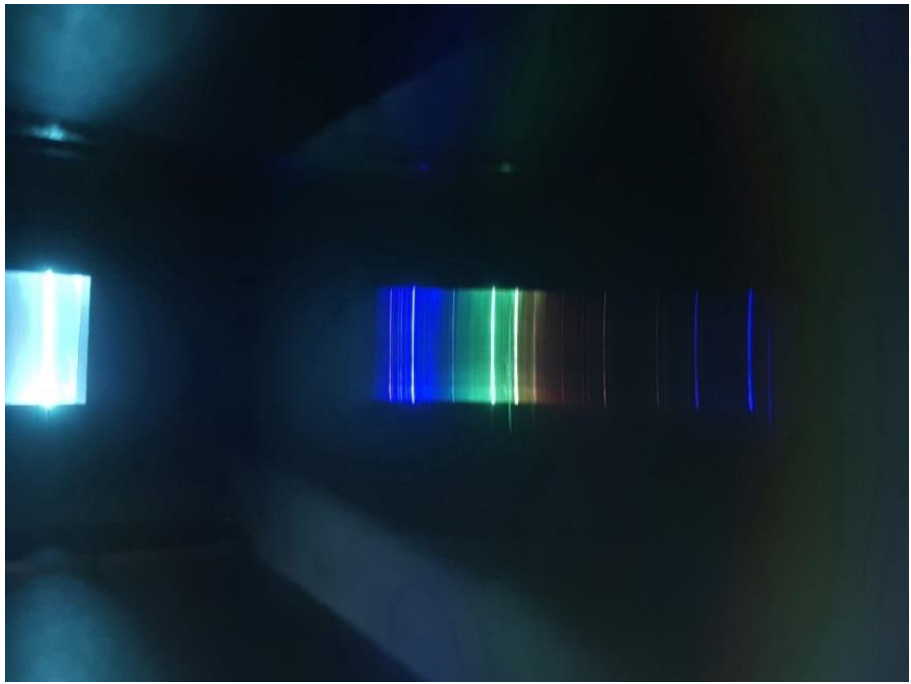

Fig. 7. The full-field-of-view image recorded by the HRPCSS includes the zeroth-order, first-order, and the second-order diffraction images of the slit.

## IV. CONCLUSION

In this paper, we demonstrated the fabrication of a smartphone spectrometer made of paper, a lab-made slit, and an inexpensive holographic transmission grating. Because the image of the slit forms the spectral image, the narrow slit made by a doctor-blade method with silver paint on a slide glass resulted in a very fine image on the image sensor of the smartphone camera, which in turn resulted in very high spectral resolution. Spectra recorded from gas discharge lamps such as Hg, Ar, and Ne using the fabricated papercraft spectrometer revealed that it had a superior resolution of 0.5 nm, which is close to that of the 3D-printed smartphone spectrometer reported in our previous study [9]. In addition, a long exposure time of the smartphone camera revealed fine and weak spectral lines not observed under short exposure times. Although the fabricated HRPCSS has much better resolution than the conventional PCSS and a resolution comparable to that of the best 3D-printed smartphone spectrometer, the build cost is less than $3, which is much lower than the cost of the smartphone spectrometers in previous studies. Therefore, the advantages of low cost, high resolution, and high sensitivity with portability make the HRPCSS a promising solution for various applications such as POC applications, disposable sensors, and science education.